# Navigate in Demanding Missions: Integrating Human Intelligence and Brain-Inspired Intelligence

Xu He, *Student member, IEEE*, Xiaolin Meng, Youdong Zhang, Lingfei Mo, *Member, IEEE*, and Wenxuan Yin

*Abstract*—This perspective analyzes the intricate interplay among neuroscience, Brain-Inspired Intelligence (BII), and Brain-Inspired Navigation (BIN), revealing a current lack of cooperative relationship between Brain-Computer Interfaces (BCIs) and BIN fields. We advocate for the integration of neuromorphic-empowered BCI into BIN, thereby bolstering the unmanned systems' reliable navigation in demanding missions, such as deep space exploration, etc. We highlight that machine intelligence, reinforced by brain-inspired artificial consciousness, can extend human intelligence, with human intelligence mediated by neuromorphic-enabled BCI acting as a safeguard in case machine intelligence failures. This study also discusses the potentials of the proposed approach to enhance unmanned systems' capabilities and facilitate the diagnostics of spatial cognition disorders, while considering associated ethical and security concerns.

*Index Terms*—Brain-inspired intelligence, brain-computer interfaces, neuromorphic, artificial consciousness

## I. INTRODUCTION

ONE primary objective in developing unmanned systems is to deploy autonomous or teleoperated machines to perform high-risk or repetitive tasks instead of humans [1]. Intelligent navigation is a core technology for various unmanned systems. However, existing navigation systems often bear limitations in environmental adaptability, robustness, and intelligence, especially in demanding missions like space and ocean exploration, polar expeditions, etc. Developing intelligent navigation technologies with controllable precision, high robustness, and ubiquitous applicability for unmanned systems is a critical and urgent scientific frontier [2]. The 2014 Nobel Prize in Physiology or Medicine was awarded to Professor O'Keefe and colleagues for their discovery of the brain's internal "GPS" and the physical basis of spatial cognition [3]. The brain's innate spatial cognition intelligence has inspired a new paradigm for intelligent navigation: Brain-Inspired Navigation (BIN).

As an emerging scientific frontier, BIN involves interdisciplinary research spanning neuroscience, Brain-Inspired Intelligence (BII), and Positioning, Navigation, and Timing (PNT), etc. Currently, a widely accepted philosophy among researchers is that BIN seeks to equip unmanned systems with brain-mimicking spatial cognition and navigation intelligence. Therefore, many efforts focus on developing brain-inspired computing methods to simulate the brain's neural mechanisms of spatial cognition and navigation, combined with intelligent perception technologies and brain-inspired hardware to develop feasible technical solutions. However, the field remains primarily in stages of theoretical exploration and partial mechanism validation, with its technical roadmap still undefined and its possibilities unbounded.

Within the domain of BII, Brain-Computer Interfaces (BCIs) have long been a significant focus, achieving notable advancements in human-computer interaction and robotic control. However, to our knowledge, no studies have yet examined the potential role of BCI in BIN research. This article aims to address this gap.

## II. A DISCUSSION ON NEUROSCIENCE, BII, AND BIN

This discussion aims to clarify the established connections among neuroscience, BII, and BIN with a simple logic, and then claim the current situation where BCI has not yet been linked with BIN.

Since its inception, AI has been profoundly influenced by neuroscience, consistently adhering to the paradigm of brain mimicry. Numerous studies have underscored the mutually beneficial relationship between AI and neuroscience, such as [4][5]. Despite AI's superman performance in visual and natural language processing, etc., it still falls short of the brain's computational efficiency and general intelligence. This disparity has driven global interest in BII, with brain-inspired computing and BCI emerging as key frontiers.

BCI, originating from the desire to augment brain intelligence, serves as an interface between the brain's nervous system and external machines. It directly uses the brain's neural signals as input, decoding these signals to enable "thought communication" and execute tasks driven by neural signals. For instance, Meta's Brain2Qwerty system [6] can decode sentences directly from brain activity using deep learning, bringing good news for patients with aphasia.

In BCI, efficient brain signal decoding remains a key technical challenge, and an increasing number of researchers

This paragraph of the first footnote will contain the date on which you submitted your paper for review, which is populated by IEEE. This work was supported by the Natural Science Foundation of Jiangsu Province under Grant No. BK20243064. Xu He receives support from the Postgraduate Research & Practice Innovation Program of Jiangsu Province under Grant No. SJCX24_0067 and SEU Innovation Capability Enhancement Plan for Doctoral Students under Grant No. CXJH_SEU 24204.

Corresponding authors: Xiaolin Meng & Lingfei Mo. (e-mail: xiaolin_meng@seu.edu.cn; lfmo@seu.edu.cn)

Authors are all with the State Key Laboratory of Comprehensive Spatiotemporal Network & Equipment Technology, School of Instrument Science and Engineering, Southeast University, Nanjing 210096, China.

Color versions of one or more of the figures in this article are available online at http://ieeexplore.ieee.org

have recognized the advantages of neuromorphic computing. This has made neuromorphic-enabled BCI decoders a promising research hotspot. For example, the team in Tsinghua University, China, developed a neuromorphic BCI decoder using memristor chips [7], optimizing decoding efficiency and reducing power consumption through hardware innovation. Additionally, with ANN2SNN technology bridging the gap between Artificial Neural Networks (ANNs) and Spiking Neural Networks (SNNs) [8], many AI-empowered BCI decoding techniques can be translated into SNN versions for neuromorphic deployment.

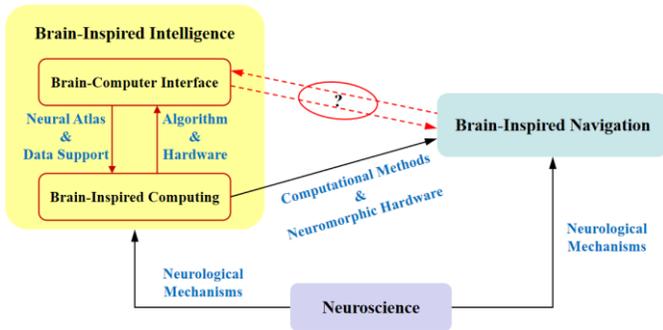

**Fig. 1** The synthesis of the relationships among neuroscience, BII, and BIN.

BIN seeks to harness the mechanisms underlying the brain's navigational intelligence from neuroscience. Utilizing the perspective of BII, it models these neural mechanisms with brain-inspired neural networks and integrates intelligent perception and neuromorphic hardware to confer spatial cognition capabilities on unmanned systems. The above pathways are akin to how today's BII seeks to replicate validated neural functions and structures through simulation, thereby approximating brain intelligence. Therefore, we believe that BIN is a vital entry point and a breakthrough for BII.

However, BCI, an evergreen topic in BII, has so far been overlooked regarding its potential connection to BIN (Figure 1). Considering that BCI can achieve "thought communication" between humans and machines by decoding human intentions, we aim to outline a potential integration pipeline for BCI and BIN. This endeavor aims to bridge the gap between human intelligence and machine intelligence to enhance unmanned systems' reliability in demanding missions, and to further strengthen the symbiotic relationship between BII and BIN.

### III. POTENTIAL PIPELINE OF THE HUMAN INTELLIGENCE-ASSISTED NAVIGATION IN DEMANDING MISSIONS

Undoubtedly, enhancing machine intelligence by imitating the brain is justified, with the ultimate aim of achieving human-like navigational capabilities. However, the non-mutually exclusive nature of human intelligence and machine intelligence is often overlooked. Machine intelligence can extend human intelligence, with human intelligence acting as a safeguard for machine intelligence failures. Therefore, we advocate for the introduction of BCI technology in BIN research to bridge human intelligence and machine intelligence, ensuring ubiquitous navigation in diverse and challenging environments.

As illustrated in Figure 2, unmanned systems encounter substantial challenges in fully autonomous operations within application scenarios such as extraterrestrial exploration, ocean exploration, polar expeditions, and low-altitude economy. We propose endowing unmanned systems with brain-inspired artificial consciousness to function as a "brain," facilitating perception, cognition, reasoning, and decision-making for PNT intelligence and maximizing autonomy and self-control. However, the absence of spatial priors, like reliable maps or cooperative conditions, poses difficulties for safe and reliable autonomous navigation during free exploration. This prompts the question: how can human intelligence serve as a fail-safe for PNT capabilities when machine intelligence falters?

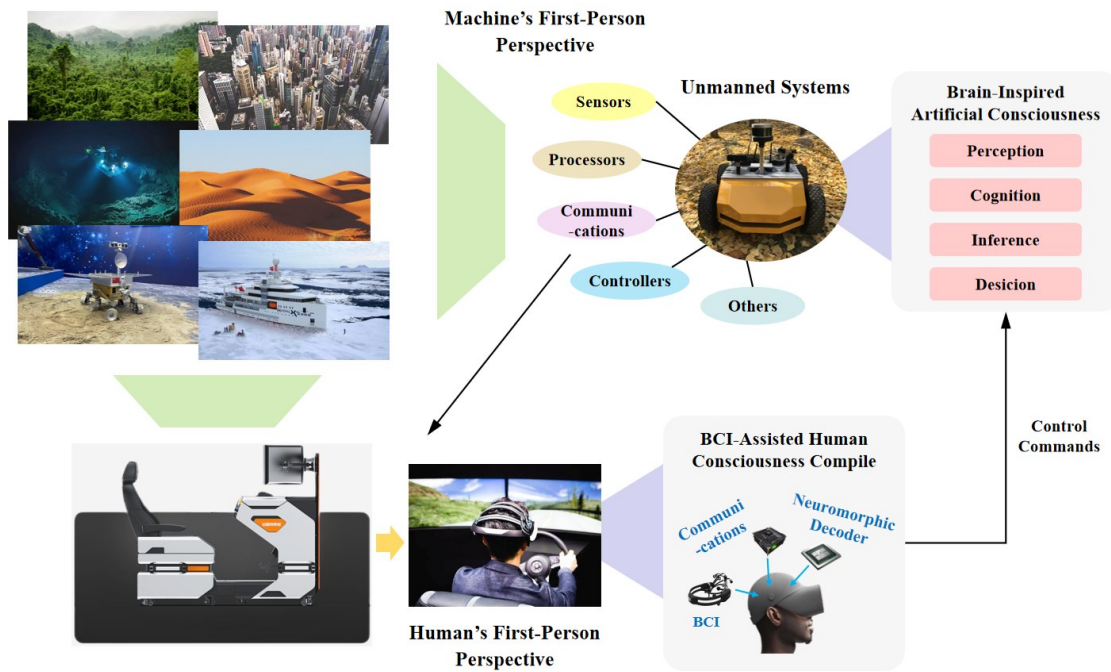

**Fig. 2** A potential integration pipeline of BCI and BIN.

Currently, some leading researchers and enterprises like Ford Global Technologies are focusing on "thought-driven driving" technology [9], which interprets human driving intentions using BCI. This pathway, facilitated by real-time communication, enables remote driving operations from a first-person perspective via consciousness, addressing the aforementioned issue. Central to the viability of thought-driven driving is BCI-assisted decoding of human consciousness. Essential infrastructures, like communication systems, are also crucial. As previously noted, neuromorphic computing has gained attention in BCI research [7]. Utilizing brain-inspired chips, this paradigm offers high energy efficiency, low latency, and low power consumption compared to von Neumann architecture-based computing units, which are advantageous for dynamic task execution of unmanned systems.

Therefore, we advocate developing integrated neuromorphic hardware and software for human intention decoding for demanding missions. By leveraging unmanned systems' communication capabilities, we can synchronize the first-person perspectives of both machine and human, receive control commands from human intelligence, and establish an intelligent PNT capability that integrates human intelligence and BII. This allows unmanned systems, operating autonomously with brain-inspired artificial consciousness, to serve as an extension of human cognition intelligence. Conversely, when the machine intelligence of BIN systems is incapacitated, neuromorphic-empowered BCI technology can utilize human intelligence as the ultimate decision-making safeguard for unmanned systems. We posit that employing brain-inspired artificial consciousness in unmanned systems to extend human cognition intelligence could benefit the medical field by offering novel BCI-based approaches for diagnosing the brain's spatial cognition disorders. Furthermore, human thought-driving data may enhance our understanding of the brain's navigation neural circuitry, fostering advancements in BIN algorithms. Lastly, we express some concerns regarding the practical implementation of the technical approaches mentioned above, primarily potential issues related to ethics and information security.